# Analysis and Enhancement of Lossless Image Compression in JPEG XL

Final Progress Report

Rustam Mamedov

Thesis CS4490Z

Department of Computer Science

Western University

March 30, 2024

Project Supervisor: Mahmoud El-Sakka, Dept. of Computer Science

Course Instructor: Nazim H. Madhavji, Dept. of Computer Science

**Glossary**
**decoder** – a program that takes in an input codestream and produces an image
**encoder** – a program that takes in an image and produces a compressed image data in the form of a codestream
**entropy encoding** – a procedure which converts a sequence of input symbols into a sequence of bits such that the average number of bits per symbol approaches the entropy of the input symbols
**entropy decoding -** procedure which recovers the sequence of symbols from the sequence of bits produced by the entropy encoder
**JPEG** – Join Photographic Experts Group
**CSV –** comma-separated values
**DCT –** discrete cosine transform
**FLIF** – Free Lossless Image Format
**FUIF –** Free Universal Image Format
**HDR** – high dynamic range
**GED –** gradient edge detection
**GAP** – gradient adjusted predictor
**MED –** median edge detection
**PNG** – Portable Network Graphics
**DPCM** – differential pulse modulation

**Structured Abstract**

**Context and motivation**
　　　　Image compression is an exciting field of research that is continuously evolving to meet the ever-increasing demand for efficient storage, transmission, and display of visual data. This research centers on a comprehensive performance analysis and enhancement of the lossless compression capabilities of JPEG XL, the most recent iteration of the JPEG family aimed at replacing its modern counterparts.

**Research question**
　　　　In what ways can the lossless component of JPEG XL be modified to increase the compression ratio, and what techniques can be applied to achieve such an increase?

**Principal ideas**
　　　　This research explores and enhances the lossless compression capabilities of JPEG XL, with a focus of increasing the compression ratio. Several key ideas and potential approaches are considered when addressing the research question. Algorithmic refinements, adaptive coding strategies, machine learning integration, and theoretical foundations from signal processing and information theory will be used and applied wherever possible to enhance performance.

**Research methodology**
　　　　This research goal aims to create new technology as well as conduct a theoretical investigation. Various tools will be used throughout the course of the research, including version control with Git, open-source image datasets such as Kodak, DIV2K, and CLIC, as well as image manipulation software like ImageMagick, Linux and Linux-based tools. In addition, the JPEG XL, WebP2, and FLIF codecs were utilized, all of which are open-source and available for download.

**Anticipated type of results**
The anticipated results from this research encompass both theoretical insights and practical advancements in the lossless compression capabilities of JPEG XL. The primary goal is to achieve higher lossless compression ratios. Anticipated results include a quantifiable improvement in the compression efficiency of JPEG XL, making it a more effective and competitive image compression standard.

**Anticipated novelty**
Anticipated innovations include adaptive entropy encoding, potential integration of machine learning for optimal compression parameters, and exploration of hybrid compression approaches, which are not part of the JPEG XL standard.

**Anticipated impact of results**
If successful, the results would help reveal more about compression theory by showcasing how state-of-the-art techniques can be combined to achieve an optimal result. The practical impact involves the improvement of the JPEG XL standard, making it more competitive and efficient, and hence more appealing to the end user.

**Limitations**
A few limitations are considered for this problem. The most significant of which is the approach of the lowest possible theoretical bound of entropy coding in modern codecs, thus making the task of increasing compression ratio furthermore limited in its scope.

Table of Contents

## 1    Introduction

As the demand for digital information continues to surge, particularly in fields such as medicine, remote sensing, and archival, the need for efficient image compression becomes increasingly important. Lossless image compression, in particular, plays a crucial role in managing the increasing volume of image data without compromising quality. Current research in image compression primarily focuses on techniques for increasing compression ratio, with emphasis being placed on novel approaches such as predictive coding, transform coding, and context modeling. Predictive coding aims to predict the value of a pixel based on its neighboring pixels, reducing redundancy in the image data. Transform coding, on the other hand, transforms the image data into a different domain where it can be more efficiently compressed. Context modeling involves analyzing the image to determine the probability of each pixel value based on its context within the image. These techniques are then typically combined with a form of parallel processing at the implementation stage, as most codecs allow for different levels of compression, commonly referred to as "effort". Higher effort often results in better compression for both lossy and lossless, however it uses more sophisticated techniques resulting in longer compression time.

The study focuses on lossless compression in JPEG XL, the newest iteration standard of the JPEG family of codecs superseding the JPEG XS standard. The study aim is twofold. Firstly, it aims to evaluate the performance of lossless JPEG XL on a diverse set of images by developing a benchmarking application capable of efficiently compressing large number of images, collecting the relevant metadata for each image, and generating a report containing the metadata. Secondly, the study seeks to analyze the algorithm behind JPEG XL lossless compression and to improve the compression ratio by introducing changes into the JPEG XL codebase, which is publicly available as open-source code. The study aims to address the question of whether it is possible to identify and use modern compression methods to enhance the lossless compression performance of the JPEG XL codec. More specifically, the study attempts to use three prediction methods with varying levels of complexity to achieve higher compression ratios. The results reveal that although on average the compression levels are below the original codec, one of the prediction methods achieves improvement for a subset of images that can be characterized by areas of smooth colour and sharp edges. The results present a novel way of improving the lossless compression performance of JPEG XL, and highlights the potential for further advancements in this area.

The investigation begins with introducing basic concepts in the realm of image compression with a focus on lossless compression. A high-level overview of the JPEG XL codec is provided to give context about the problem being addressed. The methodologies for the proposed research goals are then given. Following the methodology, the results section presents the implementation, testing, and validation details for the proposed research goals. Finally, a discussion section is presented, including threats to validity, implications, limitations and generalizability of the results obtained. A conclusion and final remarks are followed by a discussion of future work and acknowledgements.

## 2   Background and Related Work

### 2.1   Introduction to Image Compression

Digital images often contain redundant information which can be removed through compression. The process of lossless compression is reversible, meaning that once the image is compressed, it can be reconstructed back to the original image. Lossy compression on the other hand discards certain information in the image in favour of higher compression ratios. Lossless standards are typically favoured in those environments where image quality is crucial. Examples include medical imaging, aerospace, forensics, manufacturing, and many others.

### 2.2   Techniques for Image Compression

Fundamentally, lossless compression is rooted in entropy encoding and Shannon's source coding theorem. According to the theorem, the minimum number of bits per symbol needed to encode a source with a given probability distribution is equal to its entropy. The compression process generally involves two key stages: decorrelation, which reduces inter-pixel redundancy, and entropy encoding, responsible for eliminating coding redundancy. Various decorrelation techniques are employed, including prediction, transform such as discrete cosine transform (DCT) or Wavelet, and multi-resolution techniques like hierarchical interpolation and Laplacian pyramid. Entropy coding algorithms such as Huffman are frequently used in the second stage.

JPEG XL allows for both Asymmetric Numeral Systems, which tends to be more computationally efficient, as well as Huffman which is suitable for lower complexity images.

## 2.3 Evolution of Image Compression

The JPEG image compression algorithm was first introduced in 1991 in the paper "THE JPEG STILL PICTURE COMPRESSION STANDARD" (G. K. Wallace, 1991), made public by the Joint Experts Photographic Group. The aim was to create a universal compression standard for still images, for both colour and grayscale. Since then, there have been several newer standards developed by the Joint Experts Photographic Group with the intent of complementing or completely replacing the previous versions, including JPEG 2000 which exceeded the original standard in many ways (Diego Santa-Cruz et al., 2001). The latest iteration of the standard, JPEG XL, features a more powerful codec compared to its predecessors, offering better image quality and compression ratios compared to legacy JPEG (Alakuijala J. et al., 2023). Google's WebP format presents another modern approach to the problem that features both lossy and lossless compression. WebP begins by transforming an image using several different procedures, including spatial prediction that helps to reduce entropy, a colour transform that decorrelates the R, G, and B values of each pixel, and colour cache coding - a technique that allows the algorithm to use already seen image fragments to reconstruct new pixels. WebP then uses a variant of Huffman to reduce entropy on the transformed image data.

## 2.4 Overview of JPEG XL

JPEG XL is based on ideas from Google's Pik format as well as Cloudinary's Free Universal Image Format (FUIF), which itself is based on Free Lossless Image Format (FLIF). JPEG XL offers substantially better compression efficiency than existing image formats (e.g. 50% size reduction over JPEG), as well as features that are desirable for web distribution. It is a full featured codec that supports an arbitrary number of channels, and allows parallel, progressive, and partial decoding. A colour space called "XYB", which is based on the long, medium, short (LMS) colour space that describes the response of the three types of cones in the human eye, is used during the decorrelation stage (Alakuijala, J. et al., 2019). JPEG XL uses Asymmetric Numeral Systems, which is a recently developed Arithmetic Coding based entropy coder that achieves compression ratios like Arithmetic Coding but is faster during the decoding stage. On a high level, the JPEG XL codec consists of two main modes, VarDCT mode, used in lossy compression and Modular mode, utilized in lossless compression. Both modes perform similar and/or identical steps in the process of compression, but VarDCT uses additional steps to perform variable-sized DCT with adaptive quantization. The main premise of VarDCT is to divide the image into blocks (the size of the blocks may vary, hence the name) and apply DCT to the block, transforming the spatial information (pixel values) into frequency domain information (coefficients). The resulting coefficients are then quantized, resulting in loss of information, hence it is lossy. Modular mode is responsible for efficient lossless compression; however, it is also capable of performing some lossy and near-lossless functionality. It should be mentioned that JPEG XL also contains two other modes for lossless and lossy transcoded JPEG, however these are not as significant and thus are not discussed in this paper.

## 2.5 Present Research on JPEG XL

Due to the novelty of the JPEG XL, few thorough studies have been conducted on the topic of the lossless compression of the JPEG XL codec. A significant portion of the research has been focused on comparative benchmarking of the JPEG XL standard against well known formats, where JPEG XL has been shown to achieve higher compression ratios (higher is better) when tested on natural or photographic images as well as high resolution synthetic images compared to WebP and BGP (Better Portable Graphics) formats, however WebP performs slightly better on synthetic images with simple geometrical patterns (Mandeel et al., 2021). On average, lossless JPEG XL outperforms most modern lossless codecs, while supporting compression for HDR (high dynamic range) images, i.e., images with 16 bits per channel. Such impressive achievements can be largely attributed to the use of a sophisticated context model which is used in several stages of the compression process, including the prediction stage (Alakuijala et al., 2023)

## 2.6   Analysis and Research Gap

JPEG XL's lossless compression method presents a compelling solution for preserving image quality while efficiently reducing file size. However, there does exist a significant gap in the field of research of lossless compression. Specifically, investigating JPEG XL's lossless compression performance across a diverse range of image types, sizes, and complexities could reveal areas for improvement or optimization. Additionally, studying different implementations of the prediction stage and the trade-offs between higher complexity predictors and total compression time may provide useful insight into future developments of lossless compression codecs.

## 3   Research Goal and Objectives

**Goal:** The goal of this research is to increase the compression ratio or introduce other measurable improvements, such as reducing the time to encode, of the lossless component of JPEG XL. This poses significance to the field of image compression as it would demonstrate novel uses of state-of-the-art techniques and methods in conjunction with existing algorithms used by the JPEG XL standard. If the goal is achieved, it will also improve the end-user experience by reducing the compressed file size without losing quality.

**Research Question:** In what ways can the lossless component of JPEG XL be modified to increase the compression ratio, and what novel techniques, such as adaptive entropy encoding or machine learning, can be introduced to achieve this goal?

**O1:** To collect a comprehensive series of image datasets for the purposes of gathering compression metrics. The datasets shall contain a diverse set of images that vary in color intensity, noise, and edge strength, and will be non-domain-specific, such that the conclusions that are made in the following objectives are supported.

**O2:** To develop a software solution for the systematic evaluation and comparison of lossless image compression algorithms. The software will be designed to take an image dataset as input and execute lossless compression using JPEG XL. The software will present metrics such as time taken to compress, compression ratio, and other metrics in the form of plots.

**O3:** To identify potential areas of improvement for producing better compression, such as using a different color space or entropy encoding algorithm, and investigate what modifications will introduce an increase in the compression ratio, if any. Then implement said modifications in the source code and test its correctness.

**O4:** To compare and contrast, using the proposed application in **O2**, the updated JPEG XL lossless compression algorithm in **O3**. Report on the collected data and investigate the results.

## 4  Methodology

### 4.1  Lossless Compression Benchmark Approach and Development

One of the goals was to introduce an application for benchmarking various compression algorithms for the purposes of research and analysis. Thus, an application was developed to support benchmarking image compression algorithms using JPEG XL. The application was developed using Python along with compiled codec of JPEG XL, the source code for which is open source and was acquired through GitHub. Although there are programming languages that may outperform Python in speed, it proved to be sufficient for the task while carrying the benefits of fast prototyping and iterations. The benchmark was also subsequently used to analyze and compare the effectiveness of the modifications to lossless JPEG XL compression. The JPEG XL codec supports variants of arithmetic entropy encoding, also known as range encoding (Sneyers & Wuille, 2016). As the benchmark is designed to run on datasets of variable size, large datasets may cause the program to take a long time to complete. Some codecs may take longer to complete than others, for example WebP is an older compression algorithm which is typically outperformed by newer codecs like FLIF, however this is due to the nature of the codec itself rather than the benchmark. Thus, ultimately the main bottleneck is the algorithm that is used to compress and therefore the performance limitations are up to the hardware on which the program is executed. To this end, the benchmark was parallelized through multiprocessing using the internal Python concurrency library. The benchmark measures compression ratio (the ratio between the original size and compressed size), percentage of original (reciprocal of compression ratio) and time taken to compress. The measurements are made per image and the averaged quantity of each metric is reported after the process is completed. A secondary function of the benchmark is the visualization of the generated compression data. Once the compression data is saved to a CSV (comma-separated values) file, it can then be read back by the program with the compression ratios and timings visualized across the effort range, which will depend on the specific codec, using a boxplot diagram.

### 4.2  Proposed Changes to JPEG XL Prediction Approach and Development

As the JPEG XL library is written in C++, all changes that were made to the code also used C++. A form of iterative development was used to introduce new changes to the original code for the JPEG XL library. Version control with Git and GitHub was also utilized to compare and contrast different changes that were introduced to the codec. Whenever a new method or technique was implemented, it was tested in isolation before making any more changes. This approach ensured that the results measured from a given modification are accurate. The JPEG XL codebase was explored to determine potential areas of improvement for lossless compression through detailed code tracing and the use of activity diagrams. Two stages of the encoder were identified as promising candidates: feature extraction, as well as predictor and context Fmodel selection. Image features comprise of precise and dense repeated patterns as well as image noise, however, even though this stage is part of the lossless compression pipeline, it contributes very little to achieving smaller file size. Thus, the approach was to gain a understanding of the lossless pipeline and possibly improve the prediction stage Numerous prediction algorithms have been proposed over the course of developments in the area of image compression, a notable example being the gradient-adjusted predictor (GAP) introduced in the context-based, adaptive, lossless image codec (CALIC) (Xiaolin et al., 1996). Generally, all prediction algorithms work by the same intuitive principle, which is to estimate the value of the current pixel based on its context or neighboring elements, and then represent the data more efficiently by encoding the prediction error or residual. It is an innately lossless procedure, as it stores the error between the actual pixel value and the predicted pixel value, which is then used to reconstruct the image during decompression, such that it is a bit-by-bit identical copy of the original image. Two new predictors and an updated version of one of the current predictors used in JPEG XL were applied in the development process in pursuit of improvement to compression ratio, namely GAP, gradient edge detection (GED) predictor (Avramović et al., 2010), and a modified version of median edge detection (MED) predictor, which is currently used as one of the sub-predictors in JPEG XL. MED has also been used as an initial predictor in JPEG LS.

## 5   Results

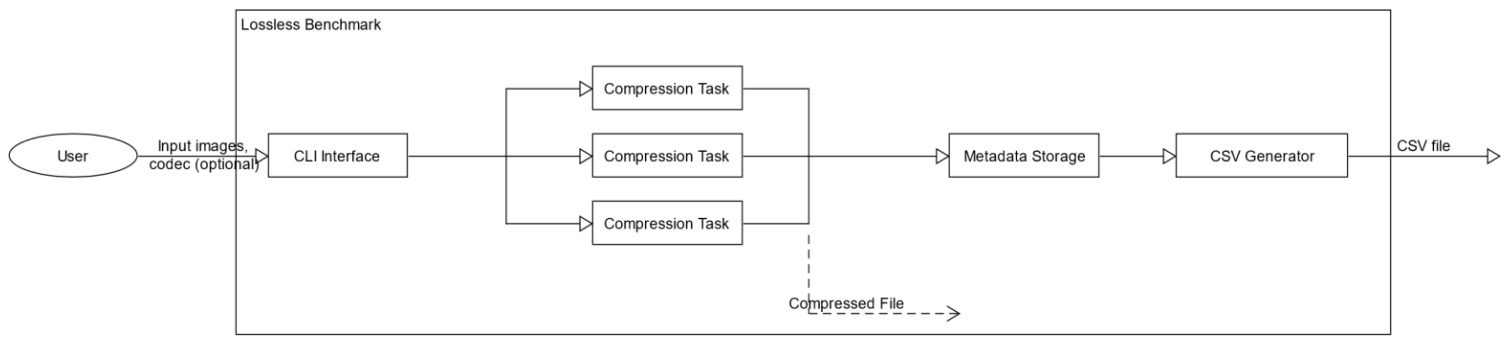

*Figure 1. Contextual diagram of the Lossless Benchmark application*

### 5.1   Lossless Compression Benchmark Key Requirements

The final version of the lossless compression benchmark was developed according to the key requirements outlined previously. The benchmark is capable of performing compression using the JPEG-XL encoder on a given target, outputting the relevant compression information into a CSV file. The benchmark features an additional command to plot the compression ratio as well as time taken to compression given the aforementioned CSV file, for visualization purposes. Requirement for parallelization was achieved using Python's internal concurrency package, where each compression task is allocated to a pool of 10 workers which was found to be optimal.

### 5.2   Lossless Compression Benchmark Design and Architecture

The program utilizes a modular architecture, with separate functions for each compression algorithm, allowing for easy expansion to include additional algorithms in the future. This was an important factor as one of the goals was to use the benchmark at a later stage to compare the compression performance of modified JPEG XL to the original. The benchmark employs a process pool to run compression tasks in parallel, optimizing the use of available resources and speeding up the benchmark process. During execution on a set of images, the application collects metadata about each image, including the filename, original file size in bytes, and image dimensions. After the compression process for a given file is complete, the application saves the relevant metadata including the compressed file size and time taken to compress the file. The percentage of original file size and compression ratio (which are reciprocals of each other) are derived and saved by the application as well. Once all tasks are complete, the benchmark dumps the collected metadata to a single CSV file. To use the benchmark, the user must specify whether to encode a set of images or to plot a reported CSV file, displayed as a boxplot. If the user proceeds with encoding, they must specify the input directory containing the uncompressed images, as well as the output directory where the compressed files will be stored. The benchmark includes several options configurable by the user, including the type of encoder to use (currently just JPEG XL), as well as the name of the aforementioned CSV file. As of right now the benchmark only accepts Portable Network Graphics (PNG) images as input, however due to the modularity of the design it can be easily extended to support other formats.

### 5.3   Lossless Compression Benchmark Implementation and Testing

The lossless compression benchmark was implemented in Python, using Git and GitHub for version control and successive iterations during the development process. Ubuntu Linux was used as the platform for developing and testing the application. Matplotlib, a plotting package, was used to create boxplot graphs for CSV file reported by the benchmark. The Typer package was utilized to collect and parse CLI arguments and flags. End-to-end testing was conducted to ensure proper functionality of the benchmark for various sizes of image sets, as well as all possible (valid) configurations of CLI arguments.

### 5.4   Lossless Compression Benchmark System Validation

Validation of the lossless compression benchmark was performed by executing the application on a test set of various images gathered in **O1,** using the full range of the effort values supported by the given encoder. The JPEG XL codec was used to test both the speed of the compression as well as the correctness of its output. The compressed files were validated against the original files by first decompressing and then comparing the two files using root mean squared error (RMSE) algorithm included as part of the ImageMagick command line tool on Linux. The CSV files generated by the application were inspected for the presence of expected row headers, and the metadata contained in the CSV files was validated against the compressed files themselves.

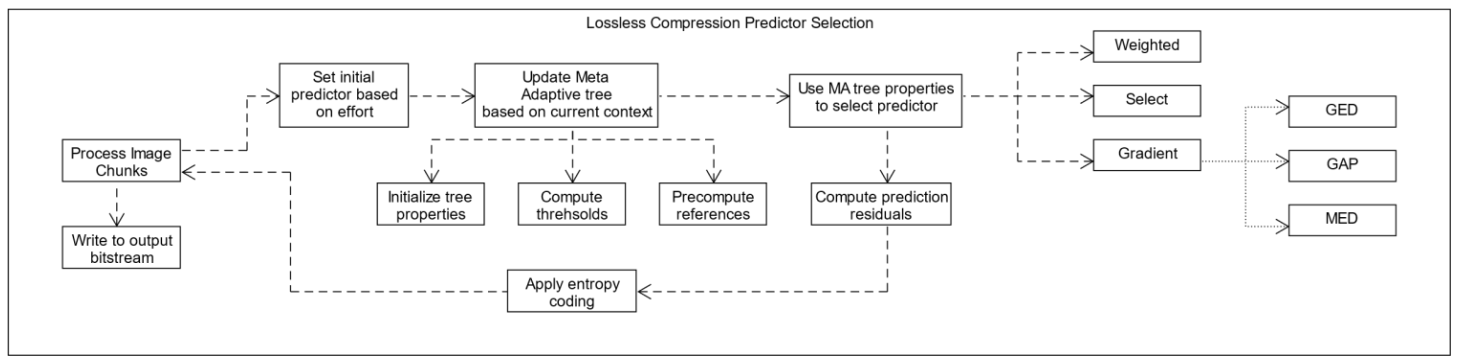

*Figure 2. Contextual diagram of proposed gradient prediction method in JPEG XL. Some of the predictors in the codec have been omitted to save space.*

## 5.5 Proposed Gradient Prediction Method Key Requirements

The successful integration of a new gradient prediction method into the JPEG XL lossless compression algorithm hinges on several key requirements. Firstly, any changes to the encoding pipeline must not break the contract between the encoder and the decoder, in other words any new changes introduced to the encoding must integrate fully with the decoding without needing to update the decoder. Secondly, the method should preserve image quality by effectively predicting pixel values without introducing artifacts or other undesirable effects. Thirdly, it should be efficient in terms of computational complexity and memory usage to ensure fast encoding and decoding speeds in order to remain practical to the end-user.

## 5.6 Proposed Gradient Prediction Method System Design and Architecture

The design and implementation of the proposed gradient prediction algorithm builds upon existing research in the realm of prediction encoding. Specifically, the two predictors which are discussed are GAP used as an initial predictor in CALIC (Xiaolin et al., 1996) and GED (Avramović et al., 2010). Additionally, a modified version of MED currently used in JPEG XL was implemented. GAP is an adaptive, nonlinear predictor that automatically adjusts itself based on intensity gradients near the predicted pixel. It outperforms linear prediction schemes such as those used in differential pulse modulation (DPCM) encoding. MED is a simpler predictor that only considers three neighboring pixels and uses the median of the three as the prediction. MED predictors are common due to their simplicity and effectiveness on most images. GED on the other hand, combines both the simplicity of MED and the efficiency of GAP into a threshold-controlled predictor (Avramović et al., 2010).

## 5.7 Proposed Gradient Prediction Method Implementation and Testing

JPEG XL includes a prediction step as the second-last stage of lossless compression before applying entropy coding to the residuals values computed by one or several predictors. In total JPEG XL utilizes an array of 16 different sub-predictors for the encoder, and 14 sub-predictors for the decoder. It is important to note that when using lossless encoding at effort value equal to 1, the algorithm will attempt to perform a "fast lossless" compression which uses single instruction multiple data (SIMD) vectorization to efficiently distribute the pixel values calculations within the processing thread pool. For the predictor implementation, all three (corrected MED, GED, and GAP) were implemented using C++ inside the libjxl codebase. The following pseudocode was used in the implementation of the corrected MED predictor:

$$
\begin{aligned}
&\text{if } C \geq max(A, B) \\
&\qquad P = max(A, B) \\
&\text{else if } C \leq min(A, B) \\
&\qquad P = min(A, B) \\
&\text{else} \\
&\qquad P = A + B - C
\end{aligned}
$$

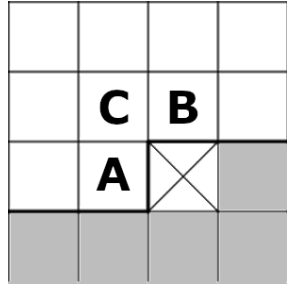

Figure 3. Neighbouring pixels for MED predictor.

GAP was implemented according to the following pseudocode as described in CALIC (Xiaolin et al., 1996):

$$g_v = |W - WW| + |N - NW| + |N - NE|$$
$$g_h = |W - NW| + |N - NN| + |NE - NNE|$$
$$if\ g_v - g_h > 80\ then\ P = W$$
$$else\ if\ g_v - g_h < -80\ then\ P = N$$
$$else$$
$$\quad P = (W + N)/2 + (NE - NW)/4$$
$$\quad if\ g_v - g_h > 32\ then\ P = (P + W)/2$$
$$\quad else\ if\ g_v - g_h > 8\ then\ P = (3P + W)/4$$
$$\quad else\ if\ g_v - g_h < -32\ then\ P = (P + N)/2$$
$$\quad else\ if\ g_v - g_h < -8\ then\ P = (3P + N)/4$$

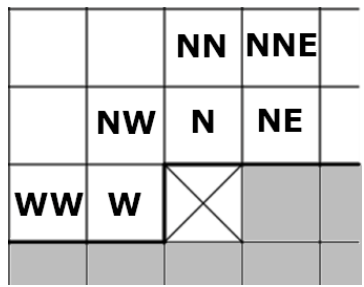

Figure 4. Neighbouring pixels for GAP predictor.

Finally, GED was implemented as follows (Avramović et al., 2010):

$$g_v = |C - A| + |E - B|$$
$$g_h = |D - A| + |C - B|$$
$$if\ g_v - g_h > T\ then\ P = A$$
$$else\ if\ g_v - g_h < -T\ then\ P = B$$
$$else\ P = 3(A + B)/8 + (C + D + E)/12$$

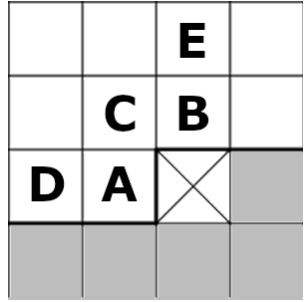

*Figure 5. Neighbouring pixels for GED predictor.*

Where *T* is a predefined threshold, which is chosen to equal 8 in the implementation. In all three implementations *P* represents the predicted pixel value.

Each implementation resided in its own separate location and was compiled from source using the CMake build system, thus resulting in three different versions of the encoder (not counting the original). Each version was then unit tested by running compression on sample images to ensure the absence of any runtime errors. Some minor issues were initially encountered that were caused by integer overflow errors on a select set of images, however the code was debugged and the errors fixed by using a different data type to store pixel values. To perform acceptance tests, the three corresponding decoders were also compiled from source and used to decompress the compressed files generated by the three modified encoders. Acceptance tests were passed after each decoder successfully decompressed the images which were exactly as the original that were passed to the encoders, without raising any error during runtime.

### 5.8    Proposed Gradient Prediction Method System Validation

Each modified versions of the encoder were executed on a total of three image data sets, each containing natural/photographic images. The results were validated and compared against the compression statistics generated by the original JPEG XL encoder. On average the three modified versions of the encoder performed worse than the original encoder if all images are considered in aggregate. However, a statistically significant improvement in compression was achieved on a select set of images from the DIV2K and CLIC data sets, that share certain characteristics, namely the images that have well defined edges as well flat areas. The GAP predictor demonstrated to be the most performant on the aforementioned set of images, followed by GED and the corrected MED.

|         | Original    | MED         | GED         | GAP         |
|---------|-------------|-------------|-------------|-------------|
| Kodak   | 459128.41   | 460495.31   | 462468.99   | 461038.46   |
| DIV2K   | 3165466.75  | 3168213.42  | 3171083.30  | 3169816.10  |
| CLIC    | 2465374.11  | 2468450.39  | 2472658.28  | 2469567.88  |

*Table 1. Average compressed file size for each predictor and image data set. File sizes are in bytes.*

| DIV2K     | CLIC                    |
|-----------|-------------------------|
| 0829.png  | larry-chen-30069        |
| 0816.png  | IMG_20150604_193209     |
| 0848.png  | milada-vigerova-7276.png |
| 0857.png  | julien-lavallee-93746.png, |
| 0893.png  | IMG_20161117_134520.png |
| 0898.png  | IMG_1194.png, 0046.png  |
| 0881.png<br>0874.png | IMG_0146.png<br>IMG_20161123_101118.png<br>IMG_20170610_124722.png<br>matthew-henry-16728.png<br>matthew-henry-16728.png<br>IMG_20170806_132019.png<br>aleksi-tappura-370.png<br>forrest-cavale-1739.png |

Table 2. List of images with sharp edges and flat areas for which improvement was achieved.

|  | MED | GED | GAP |
|---|---|---|---|
| DIV2K | 2383.60 | 1805.90 | 2990.82 |
| CLIC | 1635.96 | 1938.35 | 2605.52 |

Table 1. Average decrease in compression size for each prediction method for images containing strong edges and flat areas. Values are in bytes.

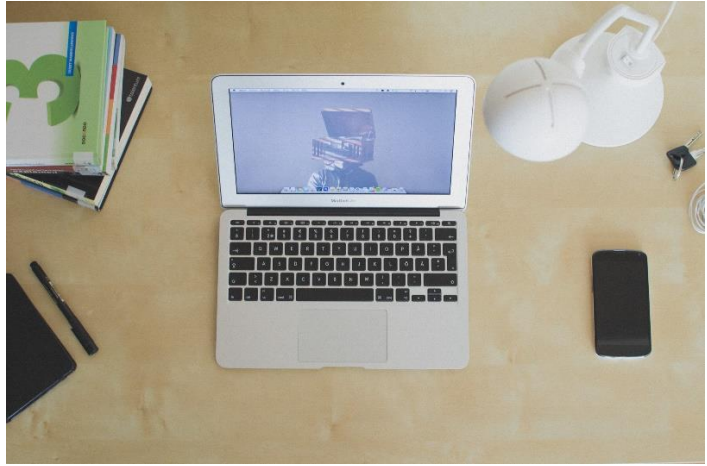

Figure 6. aleksi-tappura-370.png achieved highest reduction of 10713 bytes using the GAP predictor.

Overall, GAP performs best out of the three predictors, however MED on average outperformed GED on the DIV2K image data set. This can be explained by the relatively large frequency of high noise images within the DIV2K dataset, for which MED also serves as a median filter to reduce noise and thus achieves greater compression.

### 5.9   Novelty of Results

The results showcase a novel way of using a gradient predictor which was not previously utilized, for images containing areas of flat color and strong, well-defined edges. This method has not been previously utilized by JPEG and is therefore a good candidate for use as part of the lossless compression process.

## 6   Discussion

### 6.1   Threats to Validity

Certain components of the JPEG XL library may have an effect on the results obtained. Specifically, JPEG XL uses a meta-adaptive context model that computes the most optimal context. The context model uses a vector of integers, which are called properties to determine which context to use, which in turn determines the prediction algorithm that is used on that particular pixel. Due to the fact that the context model was not modified, it may occasionally make a suboptimal decision when choosing a predictor, reducing the compression ratio. Additionally, both GAP and GED are threshold-controlled predictors, meaning that they use pre-defined integer values to make decisions about the final prediction value. Although the thresholds were experimentally determined to be the most optimal, performance may vary depending on their value, and possibly be improved for certain images.

### 6.2 Implication of Research Results

The results present a different approach to gradient prediction in the JPEG XL lossless compression codec. The research may provide an impetus for further experimentation using improved versions of current prediction algorithms in the future. The proposed algorithm also has a practical impact in that it is able to achieve smaller file sizes for some images, which is the primary goal of compression.

### 6.3 Limitations of Results

The findings are made on a diverse set of images; however, it is nevertheless a limited pool of sample data. A larger selection of images may present more comprehensive set of results to minimize the effects of this limitation. Furthermore, the improvements become limited for images with high noise content, as GAP and GED tend to perform worse on said images compared to MED, namely because MED acts as a median filter, which is a technique used to remove image noise (Omer et al., 2018). This can also be seen by the increasing difference of compressed file sizes between MED and GAP computed on the Kodak data set with Gaussian noise applied in Table 2. Average compressed file size for each predictor, with increasing levels of Gaussian noise applied on the Kodak image data set. Values are in bytes.

|  | **MED** | **GED** | **GAP** |
|---|---|---|---|
| No noise | 460495.31 | 462468.99 | 461038.46 |
| Gaussian, variance = 0.1 | 1054747.49 | 1055819.20 | 1056822.27 |
| Gaussian, variance = 0.2 | 1182477.53 | 1184454.97 | 1186212.79 |

*Table 2. Average compressed file size for each predictor, with increasing levels of Gaussian noise applied on the Kodak image data set. Values are in bytes.*

### 6.4 Generalisability of Results

The generalisability of results obtained depends on the properties of the images being compressed, specifically if the images share similar characteristics of strong edges with flat areas, in addition to having relatively low levels of noise, similar results to the study are likely to be obtained. Generally, it is expected that more sophisticated predictors like GAP tend to outperform simpler predictors at the expense of more computation, thus when benchmarking predictors in isolation similar results are expected to occur (Karthikeyan et al., 2018).

## 7   Conclusions

The study presents a lossless compression benchmark application that allows for easy performance testing of JPEG XL compression. The application uses a distributed architecture where compression tasks for each image are allocated to a process pool of 10 workers (see Figure 1. Contextual diagram of the Lossless Benchmark application) and modularity, thus it can be extended to include support for new codecs in the future. A proposed modification to the prediction algorithm is presented, where three different predictors are compared. The gradient-adjusted prediction algorithm introduced in CALIC (Xiaolin et al., 1996) is demonstrated to outperform the median edge detection and gradient edge detection (Avramović et al., 2010) predictors when substituted for the current gradient predictor in JPEG XL. Overall, using the gradient-adjusted predictor has led to improvements for images which contain areas of flat colour along with areas of strong edges. Examples of such images include architectural, cityscapes, and nature photographs. In closing, although some improvement was achieved when using GAP, more work is required to incorporate the predictor into the context model to achieve a higher compression ratio.

## 8   Future Work and Lessons Learnt

Future work may include the investigation and optimization of the meta-adaptive context model used by the modular pipeline. Specifically, updating the context feedback loop to accommodate for the newly introduced gradient-adjusted predictor as part of the predictor selection. In addition, the modification of the fast lossless pipeline used for low effort levels to support GAP or GED instead of the current predictor, depending on the trade-off between performance and efficiency, may prove to increase compression ratios. The investigation of other gradient based prediction algorithms may reveal more insights, for example one method that was not discussed in the study is the Improved Median Edge Detection (Armin et al., 2023) algorithm, which was recently introduced and which uses a much wider range of causal pixels as well as weights similar to the weighted predictor used in JPEG XL. The lossless benchmark may be extended to support a wider range of codecs, as well as other input formats besides PNG. Moreover, additional work may be done to determine the most optimal number of workers to increase efficiently, however this number will be hardware dependent in most cases and thus can be determined experimentally.

The study reveals significant insights into lossless compression and gradient-based prediction. The study shows that incorporating gradient based prediction algorithm, namely GAP, leads to better performance on low noise images that contain smooth flat areas and areas of high edge strength. It also reveals that further experimentation on a larger sample size of data may reveal greater improvement. The study also demonstrates that using more sophisticated prediction schemes does not always guarantee better performance as was demonstrated using GAP and GED. This highlights the importance of context model feedback loop and that in order the addition of a new predictor or the modification of one necessitates the modification of the context model.

## 9   Acknowledgements

I would like to acknowledge and give my thanks to my supervisor Mahmoud El-Sakka, who has guided me through the research process and made this work possible. His advice steered me in the right direction and was crucial during all stages of the thesis. I would also like to thank a fellow student and friend, Paul Scoropan, whose work was on the lossy compression of JPEG XL and with whom I collaborated, brainstormed, and shared results with during all stages of the thesis.

**GitHub Repository Link**

https://github.com/nollum/cs4490z